\documentclass{emulateapj}

\shorttitle{Discovery of 16 new $z \sim$ 5.5 quasars}
\shortauthors{Yang et al.}

\begin{document}


\title{Discovery of 16 New $\lowercase{z}\sim5.5$ Quasars : Filling in the Redshift Gap of Quasar Color Selection}


\author{Jinyi Yang\altaffilmark{1, 2, 3}, Xiaohui Fan\altaffilmark{3, 2}, Xue-Bing Wu\altaffilmark{1, 2}, Feige Wang\altaffilmark{1, 2, 3}, Fuyan Bian\altaffilmark{4, 9}, Qian Yang\altaffilmark{1, 2}, Ian D. McGreer\altaffilmark{3}, Weimin Yi\altaffilmark{5, 6}, Linhua Jiang\altaffilmark{2}, Richard Green\altaffilmark{3}, Minghao Yue\altaffilmark{1, 3}, Shu Wang\altaffilmark{1, 2}, Zefeng Li\altaffilmark{1}, Jiani Ding\altaffilmark{3}, Simon Dye\altaffilmark{7}, Andy Lawrence\altaffilmark{8}}

\altaffiltext{1}{Department of Astronomy, School of Physics, Peking University, Beijing 100871, China}
\altaffiltext{2}{Kavli Institute for Astronomy and Astrophysics, Peking University, Beijing 100871, China}
\altaffiltext{3}{Steward Observatory, University of Arizona, 933 North Cherry Avenue, Tucson, AZ 85721, USA}
\altaffiltext{4}{Research School of Astronomy and Astrophysics, Australian National University, Weston Creek, ACT 2611, Australia}
\altaffiltext{5}{Yunnan Observatories, Chinese Academy of Sciences, Kunming 650011,China}
\altaffiltext{6}{Key Laboratory for the Structure and Evolution of Celestial Objects, Chinese Academy of Sciences, Kunming 650011,China}
\altaffiltext{7}{School of Physics and Astronomy, Nottingham University, University Park, Nottingham, NG7 2RD, UK}
\altaffiltext{8}{Institute for Astronomy, University of Edinburgh, Royal Observatory, Blackford Hill, Edinburgh, EH9 3HJ, UK}
\altaffiltext{9}{Stromlo Fellow}



\begin{abstract}
We present initial results from the first systematic survey of luminous $z\sim 5.5$ quasars. Quasars at $z \sim$ 5.5, the post-reionization epoch, are crucial tools to explore the evolution of intergalactic medium, quasar evolution and the early super-massive black hole growth. However, it has been very challenging to select quasars at redshifts 5.3 $\le z \le$ 5.7 using conventional color selections, due to their similar optical colors to late-type stars, especially M dwarfs, resulting in a glaring redshift gap in quasar redshift distributions. We develop a new selection technique for $z \sim$ 5.5 quasars based on optical, near-IR and mid-IR photometric data from Sloan Digital Sky Survey (SDSS), UKIRT InfraRed Deep Sky Surveys - Large Area Survey (ULAS), VISTA Hemisphere Survey (VHS) and Wide field Infrared Survey Explorer (WISE). From our pilot observations in SDSS-ULAS/VHS area, we have discovered 15 new quasars at 5.3 $\le z \le$ 5.7 and 6 new lower redshift quasars, with SDSS z band magnitude brighter than 20.5. Including other two $z \sim$ 5.5 quasars already published in our previous work, we now construct an uniform quasar sample at 5.3 $\le z \le$ 5.7 with 17 quasars in a $\sim$ 4800 square degree survey area. For further application in a larger survey area, we apply our selection pipeline to do a test selection by using the new wide field J band photometric data from a preliminary version of the UKIRT Hemisphere Survey (UHS). We successfully discover the first UHS selected $z \sim$ 5.5 quasar. 

\end{abstract}





\section{Introduction}
{High-redshift (z $>$ 5) quasars are important tracers to study the early Universe. However, they are difficult to be found due to both low spatial density and high contaminants from cool dwarfs when using color selection. Although more than 300,000 quasars are now known \citep[e.g.][]{schneider10,paris14,paris16}, only $\sim$ 290 quasars are at $z >$ 5. In the distribution of quasar redshift, there is an obvious gap of known quasars at 5.3 $< z <$ 5.7, due to their similar optical colors to that of late-type stars (See the redshift distribution in Sec. 4). Only $\sim$ 30 known quasars have been found in this redshift gap over a wide magnitude range (17.5 $< z <$ 26 mag) \citep[e.g.][]{stern00, cool06, romani04, douglas07, matute13, banados16}. Compared to the studies at lower redshift and higher redshift, this gap posts significant limit on the study of quasar evolution from $z \sim$ 5 to 6, over the post-reionization epoch.

Observations of the Gunn-Peterson effect using absorption spectra of high redshift quasars suggest that reionization is just completing at $z\sim$ 6, possibly with a tail to $z\sim$ 5.5 \citep{becker15, fan06, mcgreer15}. Therefore, the physical conditions of the post reionization IGM, at $z \sim$ 5-6, provides the basic boundary conditions of models of reionization, such as the evolution of IGM temperature, photon mean free path, metallicity and the impact of helium reionization \citep[e.g.][]{bolton12}. They place strong constraints on reionization topology as well as on the sources of reionization and chemical feedback by early galaxy population. Ly$\alpha$ opacity measurement directly probes the evolution of IGM. Following \cite{fan06}, several new measurements about the Ly$\alpha$ opacity at 5 $< z <$ 6 are given \citep{simpson14, becker15}, but IGM statistics are still poorly constrained at $z > 5$ \citep{becker15}. 

Moreover, a quasar sample in this redshift range is also a key to study the evolution of quasar luminosity function (QLF) and black hole growth. At high redshift, the QLF and black hole evolution have been measured at $z \sim$ 5 and 6 \citep{mcgreer13, jiang08,willott10a,willott10b,kashikawa15,yang16}. However, at $z \sim$ 5.5 they are still poorly measured due to the lack of a complete quasar sample. \cite{mcgreer13} derived the quasar spatial density at $z \sim$ 4, 4.9 and 6, and fitted a luminosity-dependent density evolution model to the combined dataset. They concluded that the quasar number density evolution steepens at high redshift, such that luminous quasars decline as a population more steeply at $z >$ 5 than from $z =$ 4 to $z =$ 5 \citep[also][]{jiang16}. However, the exact evolution of quasar density from $z$ = 5 to 6 is unclear because of the small size and high incompleteness of existing $z \sim$ 5.5 quasar sample. The quasar number density at $z \sim 5.5$ is also needed to estimate the contribution of quasars to the ionizing background just after the reionization epoch. \cite{willott10a} suggested that there was a rapid black hole mass growth phase after $z \sim$ 6. Study of black hole growth at $z \sim$ 4.8  supports that the notion of fast SMBH growth at this epoch, corresponding to probably the first such phase for most SMBH \citep{trakhtenbrot11}. Studying BH growth properties at $z\sim 5.5$ will fill in the missing link between $z\sim 5$ and 6.  

To answer the questions posted above, a large, uniformly selected sample of quasars at $5.3 < z  < 5.7$ is needed. However, so far there has not ever been a complete quasar survey at $z\sim 5.5$. As shown in \S 2, broad-band colors of $z\sim 5.5$ quasars are very similar to those of much more numerous M dwarfs, when a small number of passbands are used. Therefore, to avoid the lager number of star contaminations, previous quasar selections have always excluded the region of M dwarf locus in $r-i/i-z$ color-color diagram. As a result, most surveys of high-redshift quasars have avoided the color space occupied by $z \sim 5.5$ quasars and are highly incomplete at this redshift. To construct a large uniform $z \sim$ 5.5 quasar sample, a more effective selection to separate quasars from M dwarf in this most contaminated region is required.

In this paper, we report initial results from a new search that focuses on the selection of $z\sim 5.5$ quasars. Our new color selection criteria based on optical, near- and mid-IR colors have yielded 17 quasars in the redshift range of 5.3 $\le z \le$ 5.7 during the pilot observation described here. Our optical/IR color selection technique and candidate selection using a combination of existing and new imaging surveys are described in Section 2. The details of our spectroscopy observations and new discoveries are presented in Section 3 and Section 4. In Section 5, we discuss the completeness of our new selection and also report a test selection and first discovery using the preliminary version of the UKIRT Hemisphere Survey (UHS) photometric data. A summary is given in Section 6. In this paper, we adopt a $\Lambda$CDM cosmology with parameters $\Omega_{\Lambda}$ = 0.728, $\Omega_{m}$ = 0.272, $\Omega_{b}$ = 0.0456, and H$_{0}$ = 70 $km s^{-1} Mpc^{-1}$ \citep{komatsu09}. Photometric data from the Sloan Digital Sky Survey (SDSS) are in the SDSS photometric system \citep{Lupton99}, which is almost identical to the AB system at bright magnitudes; photometric data from IR surveys are in the Vega system. All SDSS data shown in this paper are corrected for Galactic extinction. 
}
\section{Selection of $z\sim 5.5$ Quasar Candidates}
\subsection{Using Optical and IR colors to Separate Quasars and M Dwarfs}
At  $z \sim$ 5, most quasars are undetectable in u-band and g-band because of the presence of Lyman limit systems (LLSs), which are optically thick to the continuum radiation from the quasar \citep{fan99}. Meanwhile, Lyman series absorption systems begin to dominate in the r-band and Ly $\alpha$ emission moves to the $i-$band. The $r-i$/$i-z$ color-color diagram is often used to select $z$ $\sim$ 5 quasar candidates in previous studies \citep{fan99,richards02,mcgreer13}. At higher redshift,  the $i-z$ color becomes redder and most $z >$ 5.1 quasars begin to enter the M dwarf locus in the $r-i/i-z$ color-color diagram, which makes it very difficult to select $z \gtrsim$ 5.2 quasars only with optical colors, especially at $z \sim$ 5.5, where quasars have essentially the same optical colors as M dwarfs (See Figure 1). Previous selections focused on the region in the right-bottom of the $r-i/i-z$ diagram. 

\begin{figure}[tbh]
\centering
\includegraphics[width=0.4\textwidth]{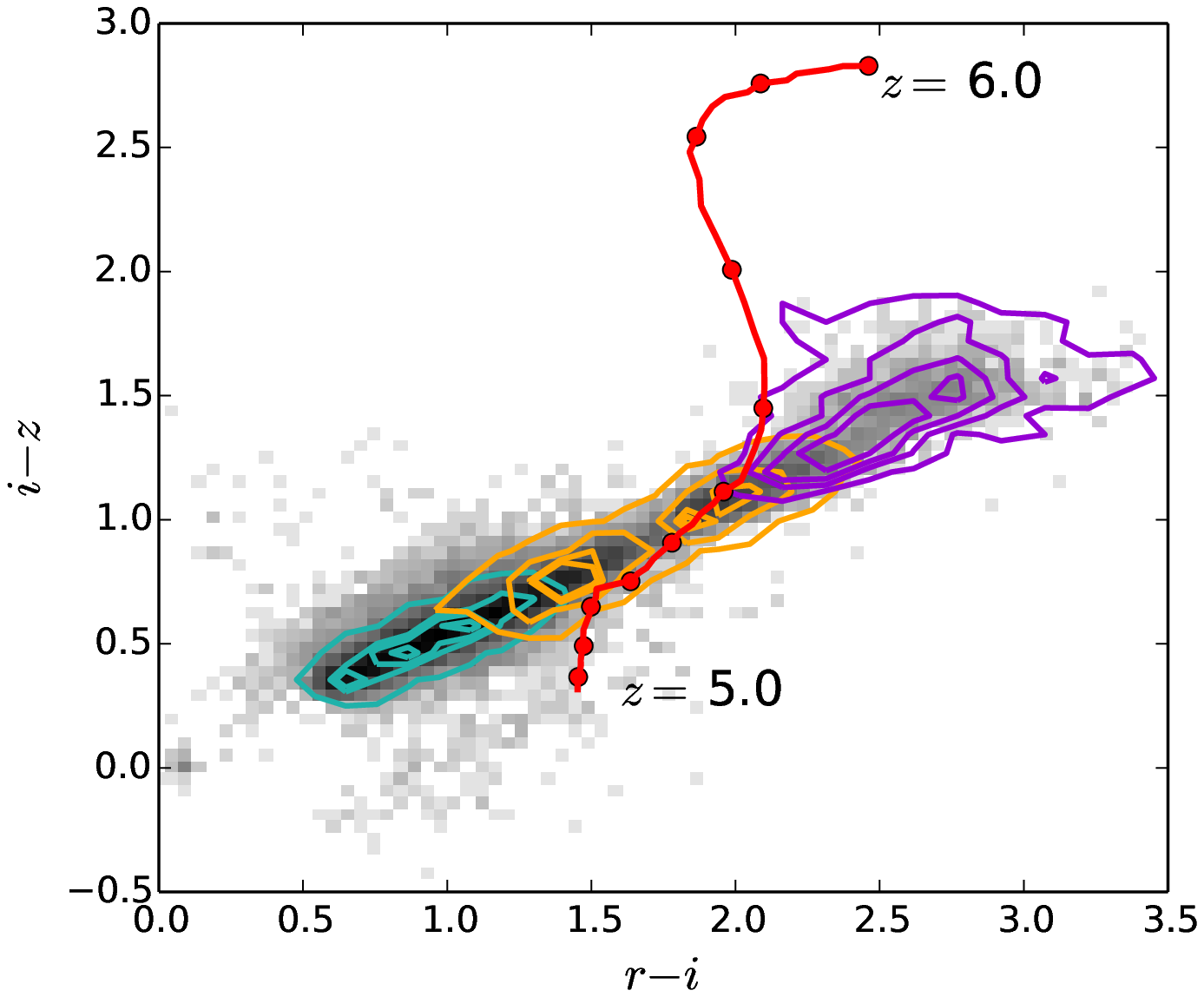}
\includegraphics[width=0.4\textwidth]{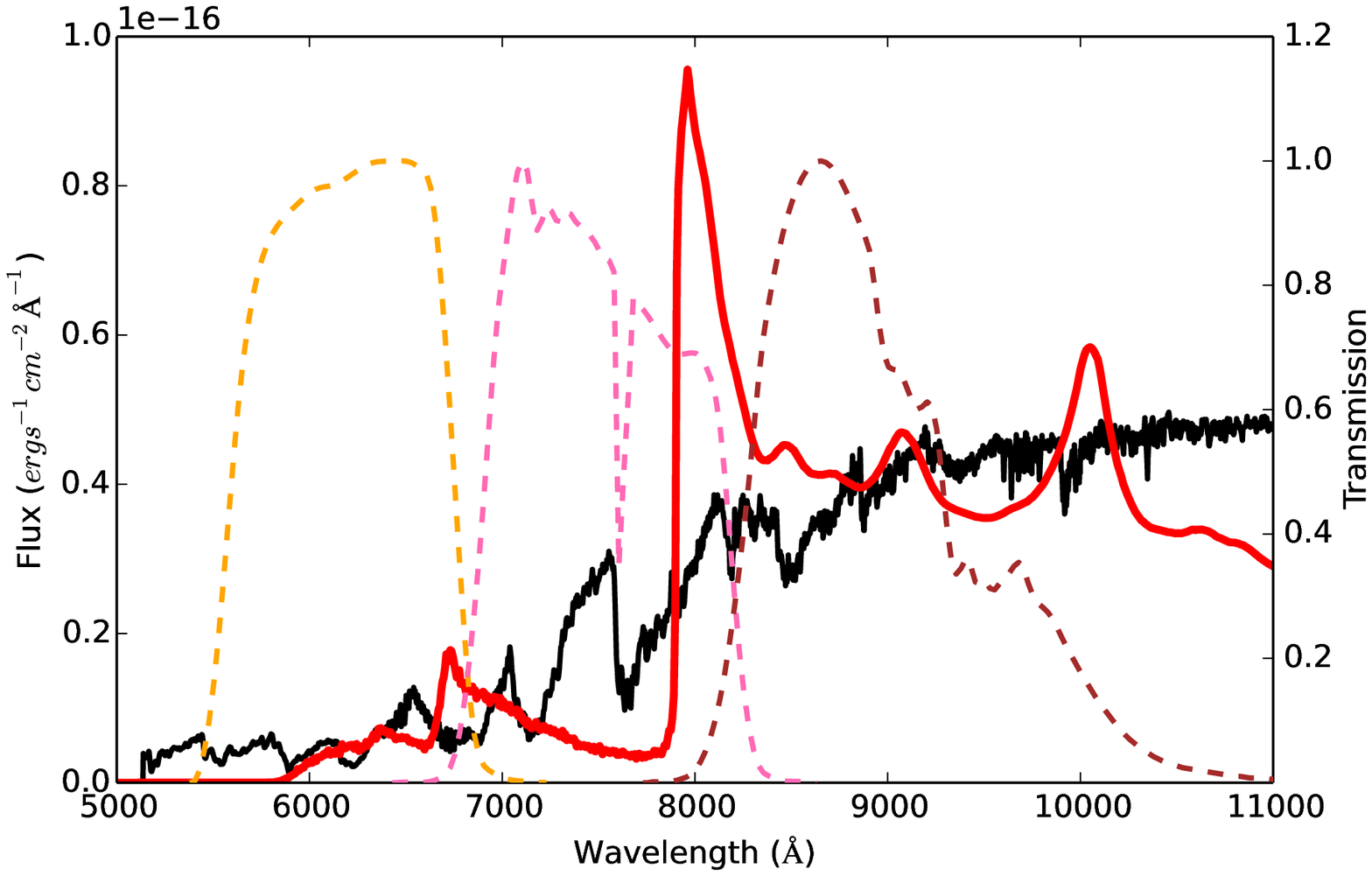}
\caption{$\bf Upper$: The color track of quasar at $z$ = 5 to 6 (Red dots and line) with a step of $\Delta z =0.1$, generated by calculating the mean colors of simulated quasars at each redshift bin.  The simulated quasar sample used here is the same sample described in Sec. 5.1. The contours show the locus of M dwarfs, from early type to late type. The cyan contours denote M1-M3 dwarfs, the orange contours denote M4-M6 dwarfs and the purple contours denote M7-M9 dwarfs. Clearly, $z\sim5.5$ quasars are serious contaminated by late type M dwarfs.
$\bf Bottom$: The spectrum of an average of simulated $z \sim$ 5.5 quasars compared with the spectrum of a typical M5 dwarf \footnote{http://dwarfarchives.org}. The dashed lines represent normalised SDSS $r$, $i$ and $z$ bandpasses from left to right. For the comparison, we scaled both the quasar spectrum and M dwarf spectrum to $i=20.0$. In this case, the synthetic SDSS $r$ and $z$ bands magnitudes of quasar are 22.1 and 18.8. For M5 dwarf, the magnitudes are 21.8 in $r$ band and 18.9 in $z$ band. It is obvious that their $r-i$ and $i-z$ colors are too similar to be distinguished using optical colors alone.}
\end{figure}

In Figure 1, we plot the quasar color track from $z$ = 5 to $z$ = 6 in the $r-i/i-z$ color-color diagram comparing with locus of M dwarfs and also show the comparison between spectra of $z \sim$ 5.5 quasar and a typical M5 dwarf. As shown, from earlier type to later type, M dwarfs show a redder $r-i$ color. M dwarfs earlier than M3 have a bluer $r-i$ color than quasars and can be more easily rejected. M dwarfs later than M4 have their continua peak at 8000-12000 $\rm \AA$ \citep{kirkpatrick93, mclean03} and can seriously contaminate quasar selection. Quasars at 5.3 $\le z \le$ 5.7 have strong Ly$\alpha$ emission line at 7600 $-$ 8150 $\rm \AA$ and a power law continuum redward of Ly$\alpha$ emission with a average index of $\alpha_{\nu}$ = $-$0.5. The locus of late type M dwarfs (M4 $-$ M8) almost overlap the whole region of $z \sim$ 5.5 quasars. 

\cite{wang16} proposed a new color selection criteria for $z \sim 5$ quasars, by adding photometric data from Wide field Infrared Survey Explorer (WISE) and at the same time relaxing the $r-i/i-z$ color cuts. ALLWISE W1-W2 color of ALLWISE can separate $z \gtrsim$ 5.1 quasars from M dwarfs due to the redder W1-W2 color of high redshift quasars than that of late-type stars. 
For quasars at $z \sim$ 5.5, if we include the whole region overlapped by M dwarf locus on the $r-i/i-z$ color-color diagram, we will include a huge number of M dwarfs. In this case, although using W1-W2 color can help to reject a part of M dwarfs, those remained M dwarfs will still result in a high contamination rate. More colors are needed to further reject M dwarfs. 

NIR photometry covering the wavelength range from 9000 $\rm \AA$ to 2$\mu$m ($J, H, K$ bands) will effectively distinguish $z\sim$ 5.5 quasars from late-type M dwarfs. The spectral energy distributions of $z \sim5.5$ quasars are mainly dominated by a power-law spectrum with a slope $\alpha_{\nu} \sim -0.5$ at the wavelength range from Ly$\alpha$ to H$\beta$, which is flatter than that of M dwarfs, a grey body spectrum. So quasars have redder $H - K$, $J-$W1 and $K-$W2 colors than that of M dwarfs. Especially, at $z\sim 5.5$, the H$\alpha$ and H$\beta$ emission lines in quasar spectra shift to W2 and W1 bands respectively. The $J-$W1, $K-$W2 and even W1-W2 colors of quasars become more redder and distinguishable form M dwarfs. We thus add $J$-W1, $H - K$ and $K -$ W2 colors, together with the $riz$ color-color diagram and W1-W2 color cut, to construct our new color selection criteria of $z\sim 5.5$ quasars.

\begin{figure*}[tbh]
\centering
\epsscale{1.2}
\plotone{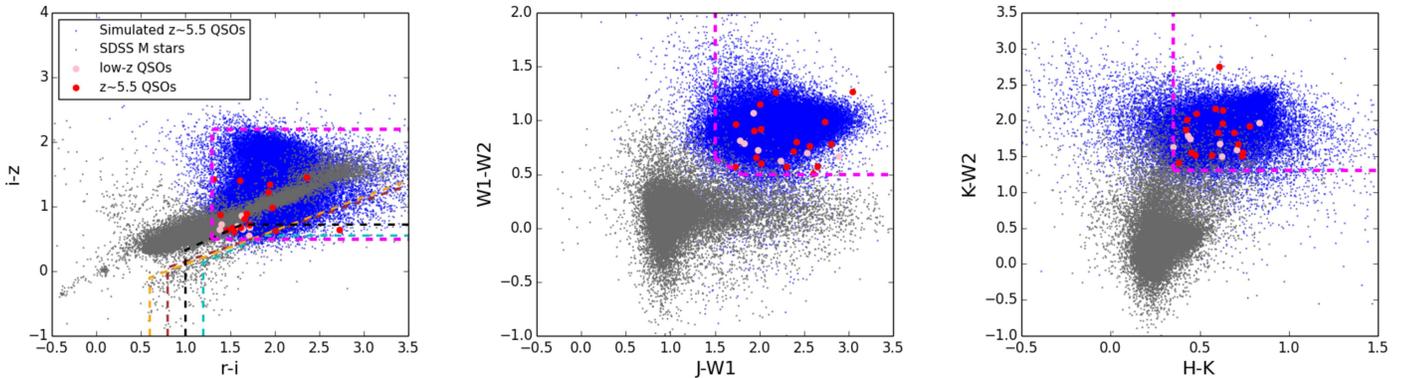}
\caption{ The color-color selection for 5.3 $\le z \le$ 5.7 quasars based on $r,i,z$,$J,H,K$ W1 and W2 bands photometric data from SDSS, ULAS and ALLWISE. We modified the traditional $r-i/i-z$ color cuts for quasars at $z \sim$ 5 to cover quasars at $z \sim$ 5.5 and add $J$, $H$, $K$ and W1 \& W2 data. Blue dots denote simulated quasars at 5.3 $\le z \le$5.7 and SDSS z band brighter than 20.8, which is the 5$\sigma$ magnitude limit of SDSS $z$ band. Grey dots show locus of SDSS Data Release 10 (DR10) spectroscopically identified M dwarfs. Our identified $z \sim$ 5.5 quasars (red solid circles), lower redshift quasars (pink solid circles) from our candidate sample are also plotted. The purple dashed lines represent our selection criteria, compared with previous $r-i/i-z$ selection pipelines for $z \sim$ 5 quasar \cite[][in brown, orange, cyan and black dashed line, respectively]{fan99, richards02, mcgreer13, wang16}. 
} 
\end{figure*}

Figure 2 shows the loci of quasars and M dwarfs in various color-color diagrams used in our new selection. For quasars, we used synthetical $r,i,z$,$J,H,K$, W1 and W2 bands photometric data from a sample of simulated quasar spectra at 5.3 $\le z \le$5.7. The details of this simulated quasar sample can be found in Sec. 5.1. We used the same photometric systems as SDSS, the UKIRT InfraRed Deep Sky Surveys (UKIDSS) - Large Area Survey (ULAS) and WISE. For M dwarfs, we used the spectroscopically identified M dwarfs from querying the SDSS DR10 catalog, and obtained their NIR photometric data from ULAS and ALLWISE. As shown, in the $riz$ color-color diagram, M dwarfs locate in a similar region as $z \sim$ 5.5 quasars. On the other hand, on the $J-$W1/W1$-$W2 and $H-K$/$K-$W2 color diagrams, most of simulated quasars could be separated from stars due to their redder $J-$W1, W1$-$W2, $H-K$ and $K-$W2 colors. To improve the efficiency of selection and reduce the size of candidate sample, we restrict our selection to the cleanest regions (purple dashed lines in Fig. 2), which are still able to include most quasars.

\subsection{Photometric datasets}
We used the following photometric datasets to select candidates of $z\sim 5.5$ quasars.
In optical range, we used SDSS DR10 photometry catalog which covered $\sim 14,400$ deg$^2$ with $u$, $g$, $r$, $i$ and $z$ bands. In near-infrared wavelengths, the published large area covered deep NIR photometric database with $Y$, $J$, $H$ and $K$($Ks$) bands are ULAS \citep{lawrence07} covering $\sim$ 4000 $deg^{2}$ in the northern sky and the VISTA Hemisphere Survey \citep[VHS;][]{mcmahon13} covering the whole southern sky. To expand the available survey area, although VHS focus on the southern sky, we also did the selection in the overlap area between VHS DR3 and SDSS, a $\sim$ 800 square degree field. For W1 and W2 bands, we used ALLWISE dataset. The ALLWISE\footnote{http://wise2.ipac.caltech.edu/docs/release/allwise/} program combined photometric data from the WISE cryogenic \citep{wright10} and NEOWISE \citep{mainzer11} post-cryogenic survey phases and mapped the entire sky with W1, W2, W3 and W4 (3.4, 4.6, 12, and 22 $\mu$m) bands. ALLWISE have high detection completeness of know quasars over almost all redshifts \citep{wang16}. Due to the shallower detections in W3 and W4 bands, we only used W1 and W2 data. 

Therefore, we carried out a quasar survey for $z \sim$ 5.5 quasars based on SDSS, ULAS/VHS and ALLWISE photometric data within a $\sim$ 4800 $deg^{2}$ field. The ALLWISE detection completeness of high redshift quasars ($z > 4.5$) is 50\% at z band magnitude $\sim$ 20.5 \citep{wang16} and decreases rapidly towards fainter end. ALLWISE W1 and W2, especially W2, are not deep enough to detect $z \sim$ 5.5 quasars with SDSS $z$ band magnitude fainter than 20.5. We thus limited our selection and main candidate sample with $z <$ 20.5 mag.

\subsection{Quasar candidate selection}
We started the candidate selection from a catalog of candidates which met our optical cuts, and selected only pointed sources. We limited our selection area to galactic latitude b $>$ 20$^{\circ}$ or b $< -20^{\circ}$ due to the quickly increasing star contaminations at lower galactic latitude. The optical color selection criteria we used for candidate selection are summarised as following, also shown in Figure 2. Here we did not limit the signal-to noise ratio ($S/N$) in $r$ band since $r$ band may be the drop-out band.

\begin{equation}
u>22.3, g > 23.3, z<20.5
\end{equation}
\begin{equation}
S/N(i) >3, S/N(z) >3
\end{equation}
\begin{equation}
g>24.0~ or ~ g-r>1.8
\end{equation}
\begin{equation}
r-i > 1.3
\end{equation}
\begin{equation}
0.5 < i-z < 2.2
\end{equation}
 
We limited $r-i/i-z$ colors and required drop-outs in $u$ and $g$ bands to ensure the redshift range of quasars. We then cross-matched optical selected objects with ULAS DR10 \& VHS DR3 and ALLWISE data using a 2$''$ cross radius. We required objects to be detected in $r$, $i$, $z$, $J$, $H$, $K$, W1 and W2 bands. We selected sources which meet our SDSS-ULAS/VHS-ALLWISE $J-$W1/W1$-$W2 and $H-K$/$K-$W2 color cuts. A $J-K$ color cut was added here for further rejection of stars. To ensure the W1 and W2 photometry quality, we limited the signal-to-noise ratio to be higher than 5 in W1 and 3 in W2. We did not further limit the $S/N$ in $J, H, K$ bands as the photometry in these bands are deep enough. Actually, all sources that met our selection had $S/N > 3$ in $J, H and K$ bands. The NIR color selections we used are listed as following (also in Figure 2).


\begin{equation}
S/N(W1) \ge 5, S/N(W2) \ge 3
\end{equation}
\begin{equation}
J-W1 > 1.5
\end{equation}
\begin{equation}
W1-W2 > 0.5
\end{equation}
\begin{equation}
W1-W2 > -0.5 \times(J - W1) + 1.4
\end{equation}
\begin{equation}
J - K > 0.8
\end{equation}
\begin{equation}
H - K > 0.35 ~ and ~ K-W2 > 1.3
\end{equation}

We visually checked images and removed targets with suspicious detections, such as multiple peaked objects or being affected by bright stars. The effect on color selection from the difference between ULAS and VHS photometry systems, especially in $K$($Ks$) band, is much smaller than photometric uncertainties. Thus we used the same color cuts for ULAS and VHS detected objects. Using this selection method, we selected $\sim$ 1000 quasar candidates in total as our main sample.

\section{Spectroscopic Identifications}
Optical spectroscopy for the identification of $z\sim$ 5.5 quasar candidates were carried out using several facilities: the 6.5m MMT telescope and the 2.3m Bok Telescope in the U.S., the 2.3m ANU telescope in Australia, the Lijiang 2.4 m telescope (LJT) in China. Our observations started in the fall of 2014. Up to date, we have observed 93 candidates from our main sample based on their brightness, colors and positions. We have discovered 21 new quasars. Three quasars, J0155+0415, J2207$-$0416 and J2225+0330 were also in our candidate list but had been observed and published earlier as $z \sim$ 5 candidates \citep{wang16}. All information of spectroscopic identifications of these 24 quasars is listed in Table 1.

We observed about 50 candidates using the Red Channel spectrograph \citep{schmidt89} on the MMT 6.5 m telescope. We used the 270 $\rm l  mm^{-1}$ grating centred at 7500 $\rm \AA$ (8500 $\rm \AA$), providing coverage from $\sim$ 5700 to 9300 $\rm \AA$ ($\sim$ 6700 to 10300 $\rm \AA$). We used  the $1\farcs0$ or $1\farcs5$ slit based on seeing condition, providing resolutions of $R\sim 640$ and $R\sim 430$, respectively. 

We also used the Wide Field Spectrograph \citep[WiFeS;][]{dopita07,dopita10}, an integral-field double-beam image-slicing spectrograph on the ANU 2.3m Telescope at Siding Spring Observatory, to observe seven of our quasar candidates. They were observed using Grating R3000 on WiFeS which gives a resolution of $R=3000$ at wavelengths between 5300$\rm \AA$ and 9800 $\rm \AA$. 

The Lijiang 2.4m telescope is located at Lijiang Observatory, Yunnan Observatories, Chinese Academy of Sciences (CAS). It is equipped with the Yunnan Faint Object Spectrograph and Camera (YFOSC) which can take spectra followed by photometric images with a very short switching time. We used Grism 5 (G5), with dispersion of 185 $\rm \AA$/mm and wavelength coverage from 5000 to 9800 $\rm \AA$. We used a $1\farcs8$ slit which yields a resolution of $R\sim550$. 

Some candidates were observed using the Boller and Chivens Spectrograph (B\&C) on Steward Observatory's 2.3m Bok Telescope at Kitt Peak with the G400 Grating and 2$\farcs$5 slit which gives a resolution of $R \sim 450$ and $\sim3400$ $\rm\AA$ wavelength coverage. 

All spectra taken on the 2.4m telescope, 2.3m Bok telescope, and MMT telescope were reduced using standard IRAF routines. The WiFeS data were reduced with a python based pipeline PyWiFeS \citep{childress14}. The flux of all spectra were calibrated using standard stars observed on the same night and then scaled to SDSS $i$-bands magnitudes for absolute flux calibration.

\begin{deluxetable*}{c c c l c c c c c}
\tabletypesize{\scriptsize}
\tablecaption{Spectroscopic information of new identified $z \sim 5.5$ quasars. \label{tbl-1}}
\tablewidth{0pt}
\tablehead{
\colhead{Name} & \colhead{Redshift} &\colhead{$z$}  &
\colhead{Instrument}  & 
\colhead{Exposure(s)}& \colhead{Grating} & \colhead{slit} & \colhead{Obsdate} 
}
\startdata
  J010806.60+071120.6 & 5.53 & 19.57 & SSO2.3m/WiFeS & 1800.0 & R3000 & 1.0 & 2014-10-15\\
  J011353.75+055951.1 & 5.00 & 20.45 & SSO2.3m/WiFeS & 1800.0 & R3000 & 1.0 & 2014-10-15\\
  J082933.10+250645.6 & 5.35 & 19.67 & MMT/Red & 300.0 & G270 & 1.0 & 2015-03-14\\
  J093523.31$-$020754.4 & 5.32 & 20.14 & MMT/Red & 250.0 & G270 & 1.5 & 2015-11-06\\
  J095712.20+101618.5 & 5.14 & 19.61 & LJT/YFOSC & 3000.0 & G5 & 1.8 & 2015-02-18\\
  J100614.61$-$031030.4 & 5.55 & 20.00 & SSO2.3m/WiFeS & 3600.0 & R3000 & 1.0 & 2015-05-15\\
  J102201.91+080122.2 & 5.30 & 19.12 & MMT/Red & 300.0 & G270 & 1.5 & 2015-11-06\\
  J113308.78+160355.7 & 5.61 & 19.71 & MMT/Red & 250.0 & G270 & 1.5 & 2015-11-06\\
  J113414.23+082853.3 & 5.69 & 20.31 & MMT/Red & 600.0 & G270 & 1.5 & 2015-03-16\\
  J114706.41$-$010958.2 & 5.31 & 19.23 & MMT/Red & 300.0 & G270 & 1.0 & 2015-05-10\\
  J114946.45+074850.6 & 5.66 & 20.30 & MMT/Red & 900.0 & G270 & 1.0 & 2015-03-13\\
  J131720.78$-$023913.0 & 5.25 & 20.08 & SSO2.3m/WiFeS & 3600.0 & R3000 & 1.0 & 2015-05-12\\
  J131929.23+151305.0 & 4.50 & 20.05 & MMT/Red & 900.0 & G270 & 1.0 & 2015-03-13\\
  J133556.24$-$032838.2 & 5.67 & 18.89 & LJT/YFOSC & 2500.0 & G5 & 1.8 & 2015-02-14\\
  J151339.64+085406.5 & 5.47 & 19.89 & MMT/Red & 600.0 & G270 & 1.0 & 2015-05-09\\
  J152712.86+064121.9 & 5.57 & 19.95 & MMT/Red & 300.0 & G270 & 1.0 & 2015-05-10\\
  J214239.27$-$012000.3 & 5.61 & 20.31 & MMT/Red & 600.0 & G270 & 1.5 & 2015-11-06\\
  J232536.64$-$055328.3 & 5.22 & 19.14 & SSO2.3m/WiFeS & 1200.0 & R3000 & 1.0 & 2015-07-20\\
  J233008.71+095743.7 & 5.30 & 19.78 & SSO2.3m/WiFeS & 1800.0 & R3000 & 1.0 & 2015-08-15\\
  J235124.31$-$045907.3 & 5.25 & 19.61 & SSO2.3m/WiFeS & 1500.0 & R3000 & 1.0 & 2015-07-20\\
  J235824.04+063437.4 & 5.32 & 19.54 & SSO2.3m/WiFeS & 1800.0 & R3000 & 1.0 & 2014-10-16\\
  \hline
  J015533.28+041506.7 & 5.37 & 19.26 & Bok/B\&C & 2400.0 & R400 & 2.5 & 2014-10-28\\
  J220710.12$-$041656.2 & 5.53 & 18.95 & LJT/YFOSC & 2400.0 & G5 & 1.8 & 2014-10-22\\
  J222514.38+033012.5 & 5.24 & 19.47 & Bok/B\&C & 2400.0 & R400 & 2.5 & 2014-10-19
  \enddata
\tablecomments{Three quasars, J0155+0415, J2207$-$0416 and J2225+0330 were also the targets in our candidate list and had been identified earlier as $z \sim$ 5 candidates \citep{wang16}. So we list them here separately.}
\end{deluxetable*}





\section{A New Sample of $z\sim 5.5$ Quasars}

From our SDSS-ULAS/VHS-ALLWISE selected candidate sample, we have observed 93 candidates and discovered 21 quasars. There are 15 new quasars in the redshift range of 5.3 $\le z \le$ 5.7. Others are lower redshift quasars but all at redshift $z >$ 5, except one broad absorption line quasar with $z = $ 4.50. There are also 3 quasars in our target list which were already observed and published as $z \sim$ 5 candidates \citep{wang16}. Two of them are z $\sim$ 5.5 quasars, the other one is at $z =$ 5.24. Therefore, in the pilot observed sample, we get a $\sim$ 25\% selection success rate for quasars and $\sim$ 18\% for the redshift range of 5.3 $< z <$ 5.7. These quasars form an uniformly selected sample of $z \sim$ 5.5 quasar with 17 quasars in the magnitude limit of SDSS $z =$ 20.5. Most of the other 72 observed candidates are M dwarfs. Few of them can only be ruled out as quasars since there are no emission features but can not be identified further.

We measure the redshifts by visually matching observed spectrum to quasar template using an eye-recognition assistant for quasar spectra software \cite[ASERA;][]{yuan13}. The matching is based on $\rm Ly \beta$, $\rm Ly \alpha$, N\,{\sc v}, O\,{\sc i}/Si\,{\sc ii}, and Si\,{\sc iv} emission lines. The typical uncertainty of our redshift measurement is around 0.03 and will be smaller for higher S/N spectra. We do not include the systematic offset of Ly$\alpha$ emission line (e.g., Shen et al.2007), which is typical $\sim$ 500 km/s and much smaller than the uncertainty of matching.

We calculate the absolute magnitude at rest-frame 1450\AA, $M_{1450}$, by fitting a power-law continuum of each observed spectrum. We assume an average quasar UV continuum slope of $\alpha_{\nu}$= $-0.5$ \citep{vandenberk01}, due to the fact that our spectra do not cover wide enough wavelength range for direct slope measurement. We normalise the power-law continuum to match the visually identified continuum windows that contain minimal contribution from quasar emission lines and sky OH lines. Those quasar spectra used for fitting are all scaled by using their SDSS $i$-band magnitude, the uncertainties of power-law continuum fitting are much smaller than the photometric errors, therefore the uncertainties of $M_{1450}$ are comparable to SDSS $i$-band photometric errors. The Redshifts, $M_{1450}$ and photometric information of our new quasars are listed in Table 2. In Figure 3, we plot the redshift distribution of our new quasars, compared with all previously known quasars and SDSS-ULAS/VHS-ALLWISE detected known quasars in the magnitude limit $z <$ 20.5 mag. Our discoveries, including the two which have been published as $z \sim$ 5 quasars, almost double the number of known quasars at $z \sim$ 5.5 with $z$ band magnitude brighter than 20.5. All spectra of new quasars are presented in Figure 4.

\begin{figure}[h]
\centering
\epsscale{1.3}
\plotone{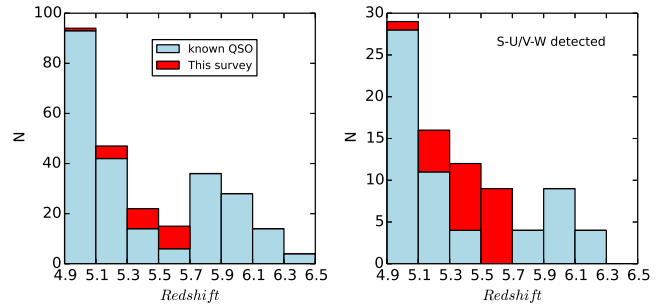}
\caption{ $Left:$ The distribution of all previously known quasars and our newly discovered quasars with $z$ band magnitude brighter than 20.5 at redshift $z >$ 4.9. Here we only count quasars within the flux limit of our survey, there are also 11 previously known $z \sim 5.5$ quasars with z band magnitude fainter than 20.5. As shown, there is an obvious gap at 5.1 $< z <$ 5.7, especially at $z \sim$ 5.5. We use the known quasar sample from the combination with the known quasar catalog in \cite{wang16} and new results in \cite{banados16}. Up to date, using SDSS-ULAS/VHS-WISE color-color selection, we have discovered 17 new quasars at 5.3 $< z <$ 5.7 and 7 lower redshift quasars, including three quasars that have been published in our $z \sim$ 5 quasar sample \citep[][also see Table 1 \& 2]{wang16}. Our optical NIR color selection is effective for finding quasars located in this redshift gap. $Right:$ The distribution of our newly discovered quasars compared with all SDSS-ULAS/VHS-WISE (S-U/V-W) detected known quasars ($z <$ 20.5 mag).
} 
\end{figure}

\begin{deluxetable*}{lllp{1.25cm}p{1.25cm}p{1.25cm}p{1.25cm}p{1.25cm}p{1.25cm}p{1.25cm}p{1.25cm}}
\centering
\tablecaption{Photometric information of new identified $z \sim 5.5$ quasars. \label{tbl-2}}
\tablewidth{0pt}
\tablehead{
\colhead{Name} & \colhead{Redshift} &\colhead{$M_{1450}$}  &
\colhead{$r$}  & \colhead{$i$}& \colhead{$z$} & \colhead{$J$} & \colhead{$H$} & \colhead{$K$} & \colhead{W1} & \colhead{W2} 
}
\startdata
J010806.60+071120.6 & 5.53 & $-$27.19  & 22.14$\pm$0.17  & 20.45$\pm$0.06  & 19.57$\pm$0.09  & 18.18$\pm$0.05  & 17.62$\pm$0.07  & 17.20$\pm$0.08  & 16.23$\pm$0.07  & 15.33$\pm$0.11\\
J011353.75+055951.1 & 5.00 & $-$25.83  & 22.72$\pm$0.22  & 21.14$\pm$0.09  & 20.45$\pm$0.17  & 19.30$\pm$0.14  & 18.35$\pm$0.13  & 17.73$\pm$0.11  & 16.76$\pm$0.09  & 16.06$\pm$0.17\\
J082933.10+250645.6 & 5.35 & $-$26.98  & 22.30$\pm$0.16  & 20.29$\pm$0.04  & 19.67$\pm$0.08  & 18.77$\pm$0.07  & 18.13$\pm$0.13  & 17.39$\pm$0.11  & 16.75$\pm$0.10  & 15.84$\pm$0.18\\
J093523.31$-$020754.4 & 5.32 & $-$26.25  & 22.61$\pm$0.21  & 20.94$\pm$0.07  & 20.14$\pm$0.14  & 19.10$\pm$0.10  & 18.47$\pm$0.11  & 18.00$\pm$0.18  & 17.14$\pm$0.14  & 16.48$\pm$0.27\\
J095712.20+101618.5 & 5.14 & $-$26.75  & 21.90$\pm$0.11  & 20.25$\pm$0.04  & 19.61$\pm$0.11  & 18.96$\pm$0.11  & 18.23$\pm$0.10  & 17.87$\pm$0.12  & 16.97$\pm$0.12  & 16.25$\pm$0.24\\
J100614.61$-$031030.4 & 5.55 & $-$26.96  & 22.96$\pm$0.38  & 20.98$\pm$0.09  & 20.00$\pm$0.14  & 18.81$\pm$0.08  & 17.94$\pm$0.07  & 17.20$\pm$0.07  & 16.20$\pm$0.06  & 15.70$\pm$0.16\\
J102201.91+080122.2 & 5.30 & $-$27.63  & 21.31$\pm$0.07  & 19.78$\pm$0.03  & 19.12$\pm$0.05  & 18.11$\pm$0.04  & 17.66$\pm$0.06  & 17.03$\pm$0.06  & 15.46$\pm$0.05  & 14.89$\pm$0.08\\
J113308.78+160355.7 & 5.61 & $-$27.49  & 22.45$\pm$0.24  & 21.13$\pm$0.10  & 19.71$\pm$0.12  & 18.74$\pm$0.09  & 18.25$\pm$0.10  & 17.53$\pm$0.10  & 16.43$\pm$0.08  & 15.87$\pm$0.15\\
J113414.23+082853.3 & 5.69 & $-$26.41  & 25.05$\pm$0.68  & 21.41$\pm$0.10  & 20.31$\pm$0.13  & 19.43$\pm$0.10  & 18.47$\pm$0.09  & 17.86$\pm$0.09  & 16.38$\pm$0.08  & 15.12$\pm$0.09\\
J114706.41$-$010958.2 & 5.31 & $-$27.44  & 21.32$\pm$0.04  & 19.86$\pm$0.03  & 19.23$\pm$0.04  & 18.00$\pm$0.07  & 17.49$\pm$0.05  & 17.10$\pm$0.07  & 16.26$\pm$0.07  & 15.70$\pm$0.14\\
J114946.45+074850.6 & 5.66 & $-$26.40  & 23.45$\pm$0.47  & 21.52$\pm$0.09  & 20.30$\pm$0.14  & 19.48$\pm$0.11  & 18.73$\pm$0.14  & 17.95$\pm$0.13  & 17.29$\pm$0.15  & 16.03$\pm$0.18\\
J131720.78$-$023913.0 & 5.25 & $-$26.27  & 22.20$\pm$0.17  & 20.80$\pm$0.07  & 20.08$\pm$0.14  & 19.42$\pm$0.16  & 18.75$\pm$0.14  & 18.30$\pm$0.18  & 17.18$\pm$0.12  & 16.56$\pm$0.28\\
J131929.23+151305.0 & 4.50 & $-$25.62  & 22.45$\pm$0.15  & 20.72$\pm$0.05  & 20.05$\pm$0.09  & 19.38$\pm$0.08  & 18.62$\pm$0.15  & 17.78$\pm$0.11  & 16.48$\pm$0.07  & 15.82$\pm$0.14\\
J133556.24$-$032838.2 & 5.67 & $-$27.76  & 22.70$\pm$0.25  & 20.34$\pm$0.05  & 18.89$\pm$0.04  & 17.76$\pm$0.03  & 17.19$\pm$0.06  & 16.49$\pm$0.05  & 15.38$\pm$0.04  & 14.67$\pm$0.06\\
J151339.64+085406.5 & 5.47 & $-$26.81  & 22.15$\pm$0.12  & 20.76$\pm$0.06  & 19.89$\pm$0.09  & 19.02$\pm$0.07  & 18.40$\pm$0.08  & 17.83$\pm$0.10  & 17.28$\pm$0.13  & 16.32$\pm$0.21\\
J152712.86+064121.9 & 5.57 & $-$26.92  & 22.96$\pm$0.28  & 21.35$\pm$0.08  & 19.95$\pm$0.10  & 18.72$\pm$0.07  & 18.11$\pm$0.08  & 17.65$\pm$0.09  & 16.70$\pm$0.08  & 16.10$\pm$0.17\\
J214239.27$-$012000.3 & 5.61 & $-$26.24  & 23.60$\pm$0.70  & 21.65$\pm$0.20  & 20.31$\pm$0.30  & 19.30$\pm$0.13  & 18.67$\pm$0.16  & 18.04$\pm$0.15  & 16.88$\pm$0.11  & 16.09$\pm$0.25\\
J232536.64$-$055328.3 & 5.22 & $-$27.13  & 21.17$\pm$0.07  & 19.78$\pm$0.03  & 19.14$\pm$0.06  & 18.17$\pm$0.06  & 17.68$\pm$0.09  & 17.05$\pm$0.09  & 16.34$\pm$0.07  & 15.55$\pm$0.12\\
J233008.71+095743.7 & 5.30 & $-$26.75  & 22.07$\pm$0.14  & 20.45$\pm$0.05  & 19.78$\pm$0.10  & 18.73$\pm$0.09  & 17.72$\pm$0.09  & 17.24$\pm$0.08  & 15.92$\pm$0.06  & 15.15$\pm$0.11\\
J235124.31$-$045907.3 & 5.25 & $-$26.34  & 22.10$\pm$0.14  & 20.47$\pm$0.05  & 19.61$\pm$0.09  & 19.14$\pm$0.17  & 18.36$\pm$0.18  & 17.92$\pm$0.21  & 17.21$\pm$0.16  & 16.14$\pm$0.24\\
J235824.04+063437.4 & 5.32 & $-$27.26  & 21.69$\pm$0.09  & 20.14$\pm$0.04  & 19.54$\pm$0.08  & 18.58$\pm$0.11  & 17.68$\pm$0.06  & 17.25$\pm$0.08  & 16.01$\pm$0.06  & 15.25$\pm$0.10\\
\hline
J015533.28+041506.7 & 5.37 & $-$27.10  & 21.70$\pm$0.10  & 19.97$\pm$0.03  & 19.26$\pm$0.06  & 18.34$\pm$0.06  & 17.62$\pm$0.06  & 17.01$\pm$0.06  & 16.33$\pm$0.07  & 15.19$\pm$0.10\\
J220710.12$-$041656.2 & 5.53 & $-$27.77  & 22.32$\pm$0.24  & 19.59$\pm$0.03  & 18.95$\pm$0.06  & 17.86$\pm$0.04  & 16.89$\pm$0.04  & 16.30$\pm$0.05  & 15.12$\pm$0.04  & 14.14$\pm$0.05\\
J222514.38+033012.5 & 5.24 & $-$27.17  & 21.74$\pm$0.14  & 20.02$\pm$0.05  & 19.47$\pm$0.10  & 18.29$\pm$0.06  & 17.99$\pm$0.14  & 17.28$\pm$0.10  & 16.50$\pm$0.08  & 15.69$\pm$0.13
 \enddata
\end{deluxetable*}

\subsection{Notes on Individual Objects}
{\bf J133556.24$-$032838.2}, $z$ = 5.67. It is one of the most luminous new quasar with $M_{1450}$ = $-27.76$. 

{\bf J152712.86+064121.9}, $z$ = 5.57. It is a weak line quasar with a very weak Ly$\alpha$ emission line and no other obvious emission features. We measure the redshift by matching the continuum to template and the redshift uncertainty is a little larger than others.

{\bf J131929.23+151305.0}, $z$ = 4.50. The quasar has the lowest redshift in our new discoveries. It is a broad absorption line quasar with redder colors than normal lower redshift quasars, that is why it could be selected by our selection.

{\bf J113414.23+082853.3}, $z$ = 5.69. This one is the highest redshift quasar in this sample with strong Ly$\alpha$ emission and strong IGM absorption blueward of Ly$\alpha$.

\begin{figure*}[h]
\centering
\epsscale{1.1}
\plotone{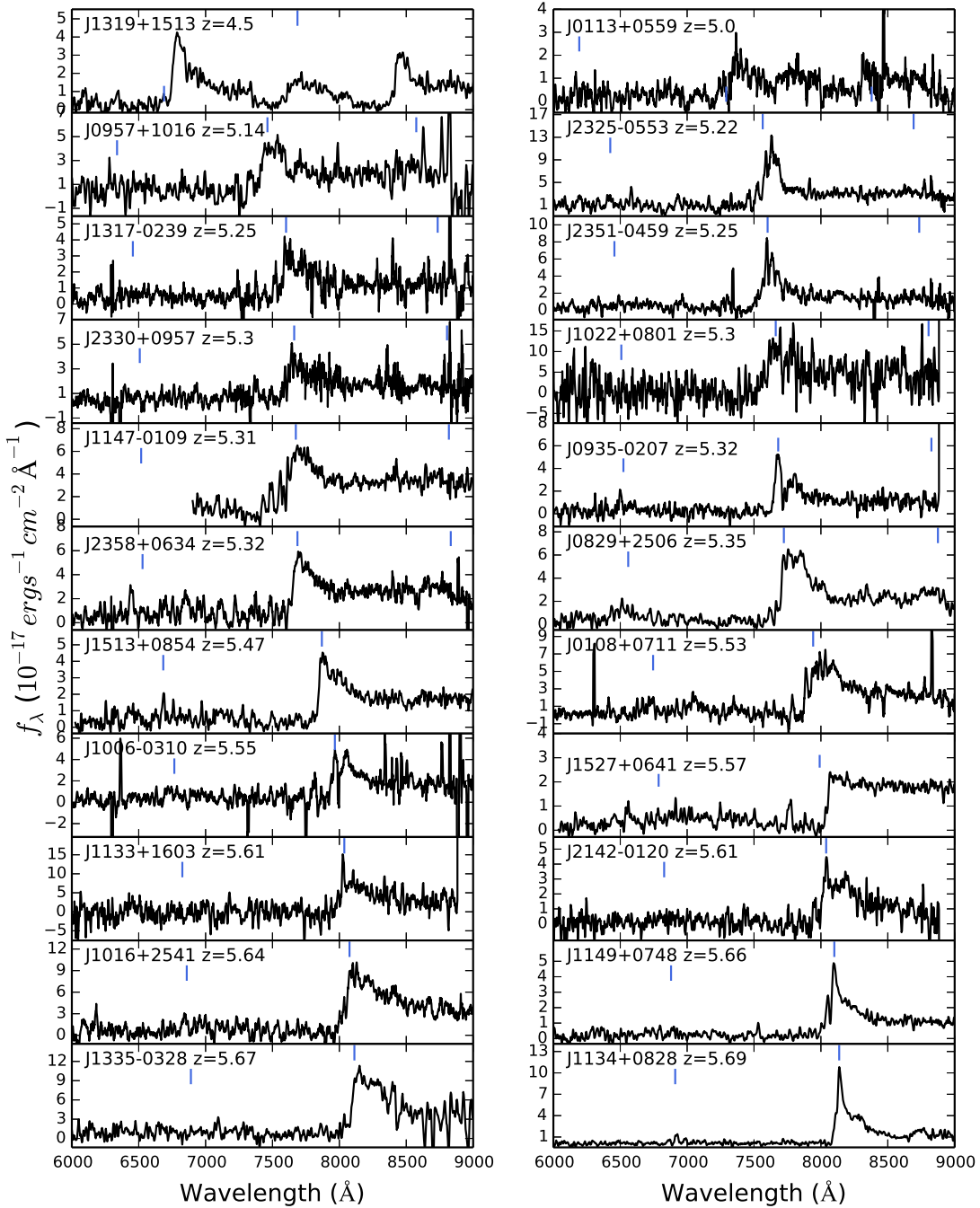}
\caption{ The spectra of our 22 new discovered quasars. 21 of them are from our main SDSS-ULAS/VHS-ALLWISE selected z $\sim $ 5.5 quasar candidate sample. J1016+2541 is from the UHS selected sample (Sec 5.1). The blue vertical lines show the $\rm Ly \beta$, $\rm Ly \alpha$ and Si\,{\sc iv} emission lines. All spectra taken using SSO 2.3m/WiFeS and P200/DBSP are smoothed with a 10 pixel boxcar. Spectra from MMT/Red and LJT/YFOSC are smoothed with a 3 pixel boxcar. All spectra are corrected for Galactic extinction using the \cite{cardelli89} Milky Way reddening law and E(B $-$ V) derived from the \cite{schlegel98} dust map.} 
\end{figure*}

\section{Discussion}
\subsection{Selection completeness}

\begin{figure}[h]
\centering
\epsscale{1.2}
\plotone{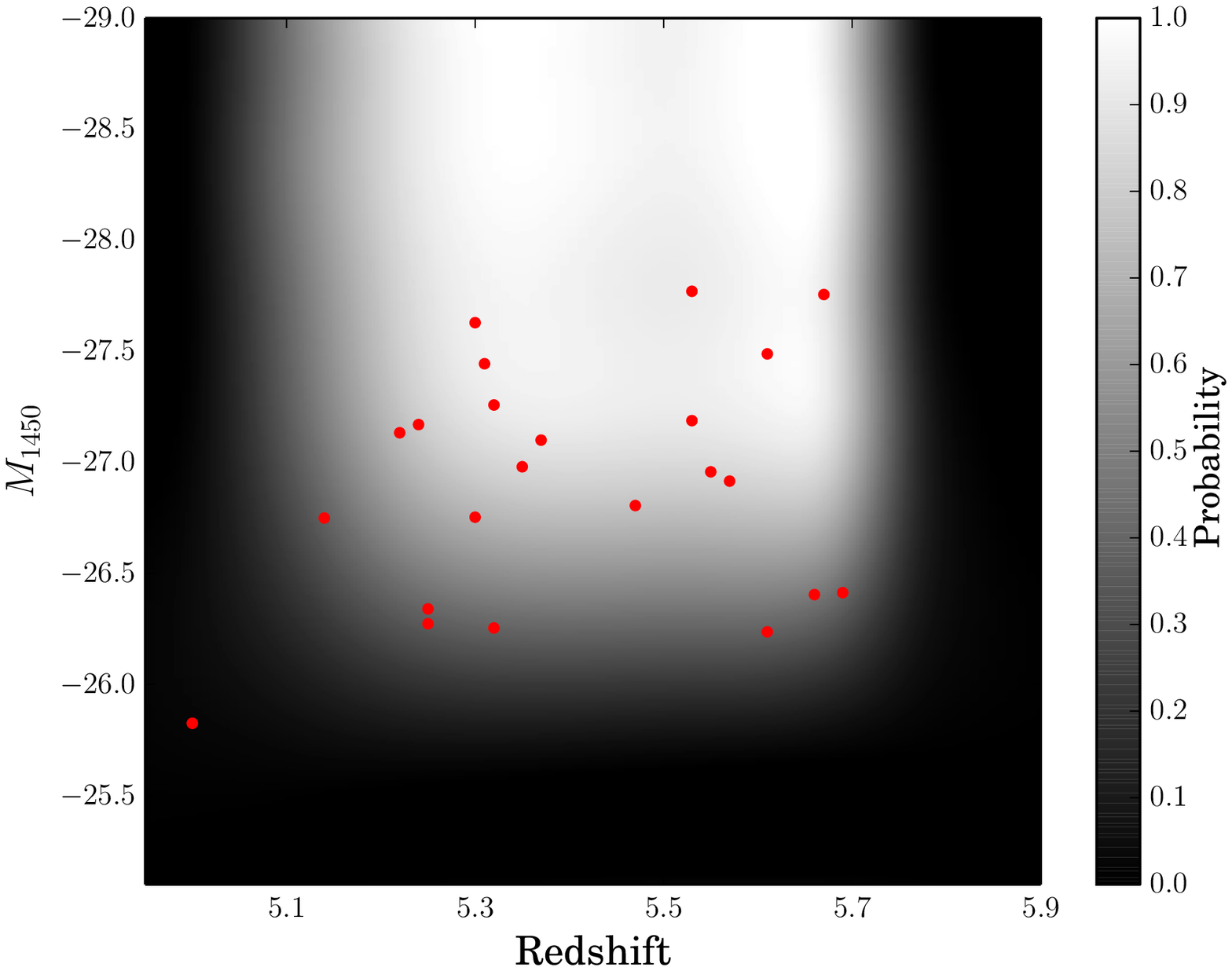}
\caption{ The selection function of SDSS-ULAS-ALLWISE color selections. Red points represent new quasars from our main candidates sample, including three quasars that have been published in our $z \sim 5$ quasar sample \citep{wang16}. The mean selection probability at 5.3 $< z <$ 5.7 is $\sim$ 91\% at $M_{1450}<$ $-$26.5 and is $\sim$ 84\% at $M_{1450} < $ $-$26. The decreasing of probability is caused by the increasing photometric uncertainties at faint end, especially in ALLWISE W2 bands. } 
\end{figure}

The sample obtained now is not yet a complete survey sample since we have only spectroscopically observed a small part of candidates, in which case it is hard to estimate the incompleteness of spectroscopy. But we want to first calculate completeness to demonstrate the effectiveness of the selection here. It will be used in later works. To first calculate the completeness of our color selection criteria, we generate a sample of simulated quasars using the quasar model from \cite{mcgreer13}. We extend this model toward redder wavelengths to cover the ALLWISE W1, W2 bands for quasars at $z$  = 5 to 6 \citep[McGreer et al. in prep.;][]{yang16}. Based on this model, a total of $\sim$ 200,000 simulated quasars have been generated and evenly distributed in the ($M_{1450}$, $z$) space of 5$< z <$6 and $-30 < M_{1450} < -25$. We assign optical photometric errors using the magnitude-error relations from the SDSS main survey. For $J,H,K$ bands, we use the ULAS photometric errors. The magnitude-error relations between ULAS and VHS are similar in the flux limit of our selection. The little differences, including the difference between ULAS K and VHS Ks will not affect the selection function too much. Here to show the redshift and magnitude dependent selection completeness, we only use the relations from ULAS photometry. We will consider a separate VHS photometry based selection function in the further works for complete sample construction and luminosity function measurement. ALLWISE detection depth is highly dependent on sky position, which will affect the detection incompleteness and photometric uncertainties. We model the coverage-dependent detection incompleteness and photometric uncertainties of ALLWISE using the ALLWISE coverage map within SDSS-ULAS/VHS area. We follow the procedure outlined in \cite{yang16}.
   
Figure 5 represents our selection function for $z\sim 5.5$ quasars. The selection function shows a high completeness in the redshift range of 5.3 $< z <$ 5.7 with a $\sim$ 91\% mean completeness at $M_{1450} < -$26.5. In the range of $-26 < M_{1450} < -26.5$, the mean selection probability is around 40\%. Towards fainter end, the completeness decreases quickly due to the increasing detection incompleteness and photometric uncertainties of ALLWISE W1 \& W2 bands. As shown, the selected region with high completeness is extended to lower redshift to $z \sim$ 5.2. That is caused by the slow evolution of quasars in $r-i$ and $i-z$ colors from $z =$ 5.1 to 5.3 (See Figure 1). To include most of $z \sim$ 5.5 quasars, we use a relative relax $riz$ cut and thus can select some lower redshift quasars, which is consistent with our result shown in Figure 3. At high redshift end, our $i-z <$ 2.2 cuts will restrict selected sample into $z < $5.7 due to the quick increasing of quasar $i-z$ color from $z =$ 5.7 to 5.8. So there is a sharp edge at $z =$ 5.7.

\subsection{SDSS - UHS - ALLWISE color selection} 
Our selection using ULAS/VHS photometric data is limited in a small area. To study the IGM and quasar number density, a larger sample is required. Therefore, the coming larger area optical/NIR surveys will provide good opportunity for $z \sim$ 5.5 quasar selection. We used data from a preliminary version of the UHS. This is a J-band survey of the northern sky (0$^\circ$ $<$ Decl. $<$ 60$^\circ$) to a depth of J = 19.6, supplementing the area already covered by UKIDSS. The survey was begun by the UKIDSS consortium, but is being completed by the new operators of UKIRT,  University of Arizona and University of Hawaii. The data is initially proprietary but is intended to be public in due course, through the same interface as UKIDSS, i.e. the WFCAM Science Archive\footnote{\url{http://surveys.roe.ac.uk/wsa/}}. The survey is briefly described by \cite{lawrence13} and will be fully reported in Dye et al. (2016 in preparation). The combination of ULAS and UHS will provides a complete J band map in the northern sky with the depth matching SDSS.

To test our selection with UHS J band, we selected a test candidate sample in the spring sky. We cross matched (2$''$ ) $riz$ selected candidates with UHS and ALLWISE catalog. We then used the same $J$-W1/W1-W2 color-cut as discussed above but without cuts related to $H$ \& $K$ bands. Due to the lack of $H$ and $K$ photometry, the fraction of M dwarf contaminations will increase, here we just focused on the bright candidates with SDSS $z <$ 19.5 to reduce the number of candidates. We observed 9 selected candidates by using Palomar P200/DBSP spectrograph and MMT/Red Channel spectrograph in April and May, 2016. 

We discovered the first UHS J band selected $z \sim$ 5.5 quasar J101637.71+254131.9\footnote{The photometric information of J1016+2541 is, $r$ = 22.53$\pm$1.05, $i$ = 20.33$\pm$0.05, $z$ = 18.96$\pm$0.05, $J$ = 17.87$\pm$0.05, W1 = 15.67$\pm$0.05 and W2 = 14.99$\pm$0.08.} at $z =$ 5.64. The spectrum was obtained using P200/DBSP with grating G316 (R $\sim$ 960 at 7500\AA), 1$\farcs$5 slit and 500s$\times$3 exposure time on April 27 \& 30 in 2016 (See spectrum in Figure 4). Data was also reduced by using standard IRAF routines. This quasar is a luminous quasar with $M_{1450}$ = -27.81. We measure the redshift and $M_{1450}$ using the same method discussed in Sec. 4.1. UHS J band photometric data is helpful for the large area $z \sim$ 5.5 quasar survey. If more NIR data will be provided, we can significantly reduce the number of candidates, such as using Pan-STARRS 1 (PS1) data \citep{chambers16}. PS1 covers the entire sky above declination $-30^{\circ}$ in the $g, r, i, z, y$ filters. The area coverage and depth, especially in the reddest narrower bands $z$ \& $y$, make PS1 to be a good choice for $z \sim$ 5.5 quasar selection. Therefore, combined PS1 and UHS data, a large uniform and completeness $z \sim$ 5.5 quasar sample can be expected. Besides, the overlap area between PS1 and VHS will add a new optical/NIR covered area for quasar selection in the Southern sky.
 
\section{Summary and Future Work}
The obvious redshift gap of known quasars at 5.3 $\le z \le $5.7 becomes a limitation of the study of IGM evolution, quasar number density and BH evolution from higher redshift to lower redshift over the post-reionaiztion epoch. This gap is caused by the same colors of $z \sim$ 5.5 quasar and late-type stars in broad optical bands. To explore quasars at this redshift range, we develop a new selection method  for $z \sim$ 5.5 quasars and build the first sample of quasars at 5.3 $\le z \le$ 5.7. Main results of our works are listed as following.

\begin{itemize}
  \item  In addition to the traditional $r-i/i-z$ color-color diagram, we add new color-color selection criteria based on ULAS/VHS $J$,$H$,$K$, and ALLWISE W1\&W2 bands. We have done a pilot survey for $z \sim$ 5.5 quasars with SDSS z band magnitude brighter than 20.5 using our new selection pipeline. 
  
  \item We have discovered 21 new quasars from our SDSS-ULAS/VHS-ALLWISE selected main candidate sample. There are 15 new quasars in the redshift range of 5.3 $\le z \le$ 5.7 and 5 quasars at redshift 5 $< z <$ 5.3. The other one is a broad absorption line quasar with $z = $ 4.50. There are also 3 quasars in our target list but already being observed as $z \sim$ 5 candidates \citep{wang16}. 2 of them are z $\sim$ 5.5 quasars. Therefore, we construct the first uniform $z \sim$ 5.5 quasar sample with 17 quasars in the magnitude limit of SDSS $z <$ 20.5. 
  
  \item The selection function shows a high completeness at $M_{1450} <  -$26, which can be expected to provide a sample of new $z \sim$ 5.5 quasars for measurement of quasar luminosity function at this redshift gap. 
  
  \item For the further application of a wide field quasar survey and to construct a larger sample, we have tried to construct the selection pipeline using UHS J band data. From our SDSS-UHS-ALLWISE selected test sample, we discovered the first UHS selected $z \sim$ 5.5 luminous quasar.
  
\end{itemize} 

In a subsequent paper, we will present the final complete sample of our $z \sim$ 5.5 quasar survey and the first measurement of quasar LF at this epoch. The evolution of quasar density from $z =$ 5 to 6 will also be constrained. Our selection focus on luminous quasars, so the new quasar sample at this redshift range provide a valuable dataset to study the IGM evolution in the tail of reionization. We will expand our selection to large survey area using new dataset such as the UHS, PS1 and the VLT Survey Telescope (VST) ATLAS \citep{shanks15}. In this work, we only use color box to select $z\sim 5.5$ quasar candidates. Recently, several modern techniques, probabilistic selections based on Bayesian model or extreme deconvolution method, have been explored to search quasars at different redshift range \citep[e.g.][]{mortlock12, dipompeo15}. The extreme deconvolution method in \cite{dipompeo15} required a sample of quasars used as training set. While at $z \sim 5.5$, there were few known quasars and most of known quasars located in the right-bottom region in $r-i/i-z$ diagram due to selection criteria. There was no good training sample representing typical colors of $z \sim 5.5$ quasars before. Our new quasar sample and selection method could provide a new training sample for future probabilistic selection. Besides, an extension of the Bayesian model from \cite{mortlock12} to include more NIR colors will also be expected to be useful for $z \sim$ 5.5 quasar selection. Additional, in future, the variability (e.g. LSST ) will also play an important role on high redshift quasar selection.

\acknowledgments
We thank the referee for providing helpful comments and suggestions. J. Yang, X.-B. Wu and F. Wang thank the supports by the NSFC grant No.11373008 and 11533001, the Strategic Priority Research Program ''The Emergence of Cosmological Structures'' of the Chinese Academy of Sciences, Grant No. XDB09000000, the National Key Basic Research Program of China 2014CB845700, and from the Ministry of Science and Technology of China under grant 2016YFA0400703. J. Yang, X. Fan and I. D. McGreer acknowledge the support from the US NSF grant AST 11-07682 and AST 15-15115.. Funding for the Lijiang 2.4m telescope is provided by Chinese Academy of Sciences and the People's Government of Yunnan Province. This research uses data obtained through the Telescope Access Program (TAP), which has been funded by the Strategic Priority Research Program "The Emergence of Cosmological Structures" (Grant No. XDB09000000), National Astronomical Observatories, Chinese Academy of Sciences, and the Special Fund for Astronomy from the Ministry of Finance in China. We acknowledge the use of the Lijiang 2.4 m telescope, the MMT 6.5 m telescope, the Bok telescope, ANU 2.3m telescope and Palomar Hale 5m telescope. Observations obtained with the Hale Telescope at Palomar Observatory were obtained as part of an agreement between the National Astronomical Observatories, Chinese Academy of Sciences, and the California Institute of Technology. This work was partially supported by the Open Project Program of the Key Laboratory of Optical Astronomy, National Astronomical Observatories, Chinese Academy of Sciences. 

We acknowledge the use of SDSS photometric data. Funding for SDSS-III has been provided by the Alfred P. Sloan Foundation, the Participating Institutions, the National Science Foundation, and the U.S. Department of Energy Office of Science. The SDSS-III Web site is http://www.sdss3.org/. SDSS-III is managed by the Astrophysical Research Consortium for the Participating Institutions of the SDSS-III Collaboration including the University of Arizona, the Brazilian Participation Group, Brookhaven National Laboratory, University of Cambridge, Carnegie Mellon University, University of Florida, the French Participation Group, the German Participation Group, Harvard University, the Instituto de Astrofisica de Canarias, the Michigan State/Notre Dame/JINA Participation Group, Johns Hopkins University, Lawrence Berkeley National Laboratory, Max Planck Institute for Astrophysics, Max Planck Institute for Extraterrestrial Physics, New Mexico State University, New York University, Ohio State University, Pennsylvania State University, University of Portsmouth, Princeton University, the Spanish Participation Group, University of Tokyo, University of Utah, Vanderbilt University, University of Virginia, University of Washington, and Yale University. This publication makes use of data products from the Wide-field Infrared Survey Explorer, which is a joint project of the University of California, Los Angeles, and the Jet Propulsion Laboratory/California Institute of Technology, and NEOWISE, which is a project of the Jet Propulsion Laboratory/California Institute of Technology. WISE and NEOWISE are funded by the National Aeronautics and Space Administration. We acknowledge the use of the UKIDSS data, and the VISTA data.

{\it Facilities:} \facility{Sloan (SDSS)}, \facility{WISE}, \facility{2.4m/YNAO (YFOSC)}, \facility{MMT (Red Channel spectrograph)}, \facility{Palomar P200/Caltech}, \facility{2.3m/ANU (WiFeS)}, \facility{Bok/Steward Observatory(B\&C)}.


\begin{thebibliography}{}

\bibitem[Ba{\~n}ados et al.(2016)]{banados16} Ba{\~n}ados, E., Venemans, B.~P., Decarli, R., et al.\ 2016, arXiv:1608.03279
\bibitem[Becker et al.(2015)]{becker15} Becker, G.~D., Bolton, J.~S., Madau, P., et al.\ 2015, \mnras, 447, 3402 
\bibitem[Bolton et al.(2012)]{bolton12} Bolton, J.~S., Becker, G.~D., Raskutti, S., et al.\ 2012, \mnras, 419, 2880 
\bibitem[Cardelli et al.(1989)]{cardelli89} Cardelli, J.~A., Clayton, G.~C., \& Mathis, J.~S.\ 1989, \apj, 345, 245
\bibitem[Childress et al.(2014)]{childress14} Childress, M.~J., Vogt, F.~P.~A., Nielsen, J., \& Sharp, R.~G.\ 2014, \apss, 349, 617 
\bibitem[Cool et al.(2006)]{cool06} Cool, R.~J., Kochanek, C.~S., Eisenstein, D.~J., et al.\ 2006, \aj, 132, 823
\bibitem[Chambers et al.(2016)]{chambers16} Chambers, K.~C., Magnier, E.~A., Metcalfe, N., et al.\ 2016, arXiv:1612.05560
\bibitem[DiPompeo et al.(2015)]{dipompeo15} DiPompeo, M.~A., Bovy, J., Myers, A.~D., \& Lang, D.\ 2015, \mnras, 452, 3124 
\bibitem[Dopita et al.(2007)]{dopita07} Dopita, M., Hart, J., McGregor, P., et al.\ 2007, \apss, 310, 255
\bibitem[Dopita et al.(2010)]{dopita10} Dopita, M., Rhee, J., Farage, C., et al.\ 2010, \apss, 327, 245 
\bibitem[Douglas et al.(2007)]{douglas07} Douglas, L.~S., Bremer, M.~N., Stanway, E.~R., \& Lehnert, M.~D.\ 2007, \mnras, 376, 1393
\bibitem[Fan et al.(2006)]{fan06} Fan, X., Carilli, C.~L., \& Keating, B.\ 2006, \araa, 44, 415 
\bibitem[Fan et al.(1999)]{fan99} Fan, X., Strauss, M.~A., Schneider, D.~P., et al.\ 1999, \aj, 118, 1 
\bibitem[Jiang et al.(2008)]{jiang08} Jiang, L., Fan, X., Annis, J., et al.\ 2008, \aj, 135, 1057
\bibitem[Jiang et al.(2016)]{jiang16} Jiang, L., McGreer, I., Fan, X., et al. 2016, \aj, Submitted 
\bibitem[Kaiser et al.(2002)]{kaiser02} Kaiser, N., Aussel, H., Burke, B.~E., et al.\ 2002, \procspie, 4836, 154 
\bibitem[Kaiser et al.(2010)]{kaiser10} Kaiser, N., Burgett, W., Chambers, K., et al.\ 2010, \procspie, 7733, 77330E
\bibitem[Kashikawa et al.(2015)]{kashikawa15} Kashikawa, N., Ishizaki, Y., Willott, C.~J., et al.\ 2015, \apj, 798, 28
\bibitem[Kirkpatrick et al.(1993)]{kirkpatrick93} Kirkpatrick, J.~D., Kelly, D.~M., Rieke, G.~H., et al.\ 1993, \apj, 402, 643  
\bibitem[Komatsu et al.(2009)]{komatsu09} Komatsu, E., Dunkley, J., Nolta, M.~R., et al.\ 2009, \apjs, 180, 330 
\bibitem[Lawrence (2013)]{lawrence13} Lawrence, A. \ 2013, Astrophysics and Space Science Proceedings, 37, 271
\bibitem[Lawrence et al.(2007)]{lawrence07} Lawrence, A., Warren, S.~J., Almaini, O., et al.\ 2007, \mnras, 379, 1599
\bibitem[Lupton et al.(1999)]{Lupton99} Lupton, R.~H., Gunn, J.~E., \& Szalay, A.~S.\ 1999, \aj, 118, 1406 
\bibitem[Mainzer et al.(2011)]{mainzer11} Mainzer, A., Bauer, J., Grav, T., et al.\ 2011, \apj, 731, 53
\bibitem[Matute et al.(2013)]{matute13} Matute, I., Masegosa, J., M{\'a}rquez, I., et al.\ 2013, \aap, 557, A78 
\bibitem[McGreer et al.(2013)]{mcgreer13} McGreer, I.~D., Jiang, L., Fan, X., et al.\ 2013, \apj, 768, 105 
\bibitem[McGreer et al.(2015)]{mcgreer15} McGreer, I.~D., Mesinger, A., \& D'Odorico, V.\ 2015, \mnras, 447, 499 
\bibitem[McLean et al.(2003)]{mclean03} McLean, I.~S., McGovern, M.~R., Burgasser, A.~J., et al.\ 2003, \apj, 596, 561
\bibitem[McMahon et al.(2013)]{mcmahon13} McMahon, R.~G., Banerji, M., Gonzalez, E., et al.\ 2013, The Messenger, 154, 35
\bibitem[Mortlock et al.(2012)]{mortlock12} Mortlock, D.~J., Patel, M., Warren, S.~J., et al.\ 2012, \mnras, 419, 390 
\bibitem[P{\^a}ris et al.(2014)]{paris14} P{\^a}ris, I., Petitjean, P., Aubourg, {\'E}., et al.\ 2014, \aap, 563, A54
\bibitem[P{\^a}ris et al.(2016)]{paris16} P{\^a}ris, I., Petitjean, P., Ross, N.~P., et al.\ 2016, arXiv:1608.06483 
\bibitem[Richards et al.(2002)]{richards02} Richards, G.~T., Fan, X., Newberg, H.~J., et al.\ 2002, \aj, 123, 2945 
\bibitem[Romani et al.(2004)]{romani04} Romani, R.~W., Sowards-Emmerd, D., Greenhill, L., \& Michelson, P.\ 2004, \apjl, 610, L9 
\bibitem[Schmidt et al.(1989)]{schmidt89} Schmidt, G.~D., Weymann, R.~J., \& Foltz, C.~B.\ 1989, \pasp, 101, 713
\bibitem[Schneider et al.(2010)]{schneider10} Schneider, D.~P., Richards, G.~T., Hall, P.~B., et al.\ 2010, \aj, 139, 2360 
\bibitem[Schlegel et al.(1998)]{schlegel98} Schlegel, D.~J., Finkbeiner, D.~P., \& Davis, M.\ 1998, \apj, 500, 525
\bibitem[Shanks et al.(2015)]{shanks15} Shanks, T., Metcalfe, N., Chehade, B., et al.\ 2015, \mnras, 451, 4238
\bibitem[Simpson et al.(2014)]{simpson14} Simpson, C., Mortlock, D., Warren, S., et al.\ 2014, \mnras, 442, 3454
\bibitem[Stern et al.(2000)]{stern00} Stern, D., Djorgovski, S.~G., Perley, R.~A., de Carvalho, R.~R., \& Wall, J.~V.\ 2000, \aj, 119, 1526 
\bibitem[Trakhtenbrot et al.(2011)]{trakhtenbrot11} Trakhtenbrot, B., Netzer, H., Lira, P., \& Shemmer, O.\ 2011, \apj, 730, 7 
\bibitem[Vanden Berk et al.(2001)]{vandenberk01} Vanden Berk, D.~E., Richards, G.~T., Bauer, A., et al.\ 2001, \aj, 122, 549 
\bibitem[Wang et al.(2016)]{wang16} Wang, F., Wu, X.-B., Fan, X., et al. \ 2016, \apj, 819, 24
\bibitem[Willott et al.(2010a)]{willott10a} Willott, C.~J., Albert, L., Arzoumanian, D., et al.\ 2010a, \aj, 140, 546 
\bibitem[Willott et al.(2010b)]{willott10b} Willott, C.~J., Delorme, P., Reyl{\'e}, C., et al.\ 2010b, \aj, 139, 906  
\bibitem[Wright et al.(2010)]{wright10} Wright, E.~L., Eisenhardt, P.~R.~M., Mainzer, A.~K., et al.\ 2010, \aj, 140, 1868 
\bibitem[Yang et al.(2016)]{yang16} Yang, J., Wang, F., Wu, X.-B. et al. \ 2016, Accepted by ApJ
\bibitem[Yuan et al.(2013)]{yuan13} Yuan, H., Zhang, H., Zhang, Y., et al.\ 2013, Astronomy and Computing, 3, 65 

\end{thebibliography}
\end{document}